\documentclass[11pt]{article}
\usepackage{psfig,amssymb,citesort,fullpage}

\def\Tr{{\rm Tr}}
\newcommand{\ket}[1]{|#1\rangle}
\newcommand{\bra}[1]{\langle#1|}

\def\m@th{\mathsurround=0pt }
\def\leftrightarrowfill{$\m@th \mathord\leftarrow \mkern-6mu
 \cleaders\hbox{$\mkern-2mu \mathord- \mkern-2mu$}\hfill
 \mkern-6mu \mathord\rightarrow$}
\def\overleftrightarrow#1{\vbox{\ialign{##\crcr
     \leftrightarrowfill\crcr\noalign{\kern-1pt\nointerlineskip}
     $\hfil\displaystyle{#1}\hfil$\crcr}}}
\begin{document}

\renewcommand{\thefootnote}{\fnsymbol{footnote}}
\begin{titlepage}
\begin{flushright}
UFIFT-HEP-00-9\\
hep-th/0004129
\end{flushright}

\vskip 1.5cm

\begin{center}
\begin{Large}
{\bf String from Large N Gauge Fields\\ via Graph 
Summation 
on a $P^+$ - $x^+$ Lattice\footnote{Talk given at the {\it Fifth Workshop on
QCD}, held at Villefranche-sur-Mer, France, 3-7 January 2000.}\footnote{Supported in part by
the Department of Energy under grant DE-FG02-97ER-41029}}
\end{Large}

\vskip 2.cm

{\large Charles B. Thorn\footnote{E-mail  address: thorn@phys.ufl.edu}}

\vskip 0.5cm

{\it Institute for Fundamental Theory\\
Department of Physics, University of Florida,
Gainesville, FL 32611}

(\today)

\vskip .5cm
\end{center}

\begin{abstract}
\noindent I describe renewed
efforts to establish a string description of large $N_c$
QCD by summing large ``fishnet'' diagrams. 
Earlier work on fishnets indicated that
the usual relativistic (zero thickness) string theory 
can arise at strong 't Hooft coupling, at best yielding a 
highly idealized description, which fails to
incorporate such salient features of continuum QCD as asymptotic freedom and
point-like constituents. 
The recently conjectured AdS/CFT correspondence is compatible
with such limitations because it also gives a simple picture 
of large $N_c$ gauge theory only at strong coupling. 
In order to better understand how string theory could
emerge from large $N_c$ QCD at strong coupling, Klaus Bering, Joel Rozowsky,
and I have developed an improved implementation of my effort
of the late seventies to digitize the planar diagrams of large $N_c$
light-cone quantized QCD by discretizing both $P^+$ and
$x^+$. This discretization allows
a strong coupling limit of the sum of planar diagrams 
to be defined and studied. It also provides
a natural framework to explore the possible dual relationship
between QCD in light-cone gauge and string theory quantized on the light-cone.
\end{abstract}

\end{titlepage}

\section{Introduction}
It has long been thought that 't Hooft's $N_c\to\infty$ 
limit \cite{thooftlargen} of $SU(N_c)$
gauge theory might be usefully described by some sort of string theory.
However, there is an apparently devastating argument, that this ``QCD String''
(a.k.a. a tower of glueballs) 
is {\it not} fundamental string (of ``string theory''): the graviton
appears in the spectrum of the latter, contradicting the well-known
folk theorem \cite{weinbergw} forbidding massless spin 2 bound
states in a Poincar\'e invariant quantum field theory
\footnote{Incidentally,
the conceptual problems with the formulation of string theory (and
its low energy limit, quantum gravity) would disappear if string
(including its graviton) were a composite
structure arising in ordinary flat space quantum field theory.
Sakharov proposed this idea for quantum gravity long ago \cite{sakharov}, and
it was later vigorously explored by Adler \cite{adlerinducedgrav} and 
Zee \cite{zeeinducedgrav}. Early in this game, it was realized that the
underlying flat space theory cannot be Poincar\'e invariant,
because of this same folk theorem.
My suggestion that string be a composite of string bits \cite{thornmosc} 
also evades the theorem because the bits live in at least one less 
space dimension than string, so the space-time symmetry of the 
underlying string bit dynamics is 
only a subgroup of the Poincar\'e group of string.}. 

A way to evade this argument has been shown by the conjectured equivalence
between classical IIB superstring theory on an AdS$_5$ background and 
${\cal N}=4$ supersymmetric $SU(\infty)$ Yang-Mills on flat
4 dimensional Minkowski space-time \cite{maldacena,gubserkp,wittenholog}.
The point is that in this example the graviton lives in
5 space-time dimensions and the flat space-time global
symmetry (Poincar\'e(3,1)) is only a subgroup of Poincar\'e(4,1),
which is realized locally, not globally.
Thus the ``massless'' 5 dimensional graviton 
is a composite of the quanta of a flat-space quantum field theory
in 4 dimensional space-time. There is no massless
spin 2 particle in this 4d quantum field theory and no folk theorems
are violated. In the ${\cal N}=4$ conformally
invariant example, the projection of the 5d graviton 
onto the 4d Minkowski boundary of AdS$_5$ is a multi-gluon continuum state. 
But if the mechanism can be extended to the non-conformally invariant
gauge theory describing the gluon sector of QCD, the graviton 4d remnant would
presumably be a massive spin 2 glueball.

To move these statements beyond conjecture, one clearly needs to
establish the ``dual'' description starting from either of the
supposedly equivalent theories. I think it is clear that the 
best starting point for such a project is the flat space
quantum field theory. Unlike the previous conjectured dualities in string
theory, which asserted the equivalence of pairs of poorly defined theories,
this duality asserts the equivalence of a poorly defined theory
(string or quantum gravity) to a perfectly well defined theory
(asymptotically free or conformally invariant quantum gauge
theory on flat 4d space-time). Indeed, I am inclined to regard
this ``duality'' as more analogous to the alternate descriptions
of superconductivity given by BCS theory on the one-hand and
Landau-Ginzburg theory on the other. If this metaphor holds,
the flat space quantum field theory should be embraced as the
long sought microscopic formulation of string/quantum gravity.
 
As 't Hooft showed in his pioneering paper \cite{thooftlargen}, the
$N_c\to\infty$ limit of $SU(N_c)$ Yang-Mills theory reduces to
a certain sum of planar Feynman diagrams. Elegant techniques,
involving the explicit elimination of the off-diagonal matrix elements
of the matrix field,
have been used to obtain this limit in matrix theories of extremely
low space-time dimension (namely D=0,1) \cite{brezinipz}, but
these methods have failed to deal with theories with space-time
dimension $D>1$. At the moment,
I see no better approach to the 
$D>1$ case than
setting up a framework to carry out the direct summation of planar graphs.
In the mid-1970's, motivated by the success of light-cone quantization
of string theory \cite{goddardgrt}, I proposed \cite{thornfishnet} that planar
diagram sums be carried out by using light-cone parameterization
$x^{\pm}\equiv(t\pm z)/\sqrt2$ and that a convenient way to digitize
the sum was to discretize the momentum conjugate to $x^-$, $P^+=lm$
with $l=1,2,\cdots$,
and imaginary light-cone time $ix^+=ka$, with $k=1,2,\cdots$.
In those first papers I restricted attention to scalar field theories,
but Brower, Giles, and I soon made a first attempt to extend
the approach to QCD \cite{browergt}. 
In our setup, the strong 't Hooft coupling limit
$N_cg^2\to\infty$ favors the fishnet \cite{nielsenfishnet} diagrams
that lead to a light-cone string interpretation. Of course, by its
very nature a strong coupling limit probes the microscopic details of
discretization, and can at best show only rough qualitative resemblance to
the continuum theory. 
Even so, there were a number of loose ends and unsatisfactory features of
this first discretization of QCD which needed to be addressed.

Motivated by the goal of discovering
a more definitive string description of large $N_c$ QCD
Klaus Bering, Joel Rozowsky, and I set out to remedy these
shortcomings, and in this talk I would like to tell you about the
results of our efforts \cite{beringrt}. As you will see, we have obtained
a much improved discretization setup, but have just begun to
explore its usefulness in capturing a string picture of the
sum of planar diagrams.
\section{String on a Light-Front}
\label{lcstring}
\setcounter{equation}{0}
An evolving string sweeps out a world sheet $x^\mu(\sigma^1,\sigma^0)$
in space-time. One can choose the parameters so that $x^+=\sigma^0$,
and so that ${\cal P}^+$, the density of $P^+$, is a constant $T_0$.
Then evolution in $x^+$ is generated by the Hamiltonian
\begin{eqnarray}
H\equiv P^-=\int_0^{P^+/T_0}d\sigma\left[{{\bf\cal P}^2\over2T_0}
+{T_0\over2}{\bf x}^{\prime2}\right],
\end{eqnarray}
where for brevity we have called $\sigma^1=\sigma$ and the prime denotes 
differentiation with respect to $\sigma$. Of course ${\bf\cal P}(\sigma)$ is
the momentum conjugate to ${\bf x(\sigma)}$. A key novelty here is that
$P^+/T_0$ measures the quantity of string present, and its interpretation
as a component of momentum is secondary and derivative. In this way
the string is seen to enjoy a Galilei invariant dynamics as it moves
only in the $d-2$ dimensional transverse space.

As shown by Mandelstam \cite{mandelstam}, interactions are easily introduced
by using the path history form of quantum mechanics and including
histories in which strings break and join. Technically this is
accomplished by first obtaining the imaginary time $ix^+\equiv\tau$ 
path integral representation of $\bra{f}e^{-\beta H}\ket{i}$
for free string. The action in this case is just
\begin{eqnarray}
S^{Free}(\beta)={T_0\over2}
\int_0^\beta d\tau\int_0^{P^+/T_0}d\sigma\left[{\bf\dot x}^2
+{\bf x}^{\prime2}\right].
\end{eqnarray}
Here the action is seen to be an integral over a simply connected 
rectangular domain for an open free string and a cylinder for
a closed string. A diagram describing an arbitrary number of splits and joins
is obtained by allowing some number of cuts, each at constant $\sigma$ 
but of varying length, within the domain. The ends of these
cuts mark the splitting or joining points, and  ${\bf x}$
is discontinuous across them. Calling the generic such domain
$\Sigma$, the complete amplitude is then given by
\begin{equation}
{\cal M}=\sum_\Sigma\int{\cal D}{\bf x}e^{-S(\beta, \Sigma)}.
\end{equation} 
\section{Discretization}
\label{discrete}
\setcounter{equation}{0}
In order to give a nonperturbative definition of the path
integrals appearing in the previous section, Roscoe Giles and I
introduced a lattice version \cite{gilest} of the domains $\Sigma$. It was only
necessary to discretize $\tau$ and $\sigma$. So we set $\beta=(N+1)a$
and $P^+=Mm\equiv MaT_0$, with $M$ and $N$ fixed positive integers.
Then ${\bf x}(\sigma,\tau)$ is replaced by ${\bf x}_{lk}$, and the
functional integration by ordinary integrals. Finally, the action
is simply replaced by 
\begin{equation}
S={T_0\over2}\sum_{L\in\Sigma}\Delta{\bf x}_L^2,
\end{equation}
where $L$ labels a link on the lattice. Links between nearest
neighbor sites in the $\tau$ direction are all present, but
those in the $\sigma$ direction are occasionally absent reflecting
the possibility of splits and joins. It is a highly nontrivial
fact that this apparently noncovariant setup turns out, after
the continuum limit, to be fully compatible with Poincar\'e
invariance in the critical dimension.
\section{Feynman Diagrams on a Light-Front}
\label{lcdiagrams}
\setcounter{equation}{0}
To understand how Feynman diagrams look in light-cone
parameterization, consider the mixed representation of
the scalar field propagator:
\begin{equation}
\Delta({\bf p},p^+,ix^+)=\theta(x^+p^+){e^{-ix^+({\bf p}^2+\mu^2)/2p^+}
\over2|p^+|}.
\end{equation}
For simplicity we may establish the convention that each line
propagates forward in $x^+$, and correspondingly $p^+>0$. Also we
may pass to imaginary time, $\tau=ix^+>0$ and then write 
\begin{equation}
\Delta({\bf p},p^+,\tau)={e^{-\tau({\bf p}^2+\mu^2)/2p^+}
\over2p^+},\qquad{\tilde\Delta}({\bf x},p^+,\tau)
=\left({p^+\over2\pi\tau}\right)^{d/2}{e^{-p^+{\bf x}^2/2\tau-\mu^2\tau/2p^+}
\over2p^+}
\end{equation}
where the second form is in transverse coordinate representation and
$d=D-2$ is the dimensionality of transverse space.

Discretizing $\tau=ka$ and $p^+=lm=laT_0$, $k,l=1,2,\cdots$,
the coordinate propagator becomes\footnote{The discretization of
$\tau$ provides a universal ultraviolet cutoff, and every
diagram will therefore be finite. In contrast, the DLCQ industry
\cite{brodskyppreport} 
keeps time continuous, and must regulate ultraviolet divergences in
some other fashion.}
\begin{equation}
{\tilde\Delta}_{lk}({\bf x})
=\left({lT_0\over2k\pi}\right)^{d/2}{e^{-T_0(l/2k){\bf x}^2+k\mu^2/2lT_0}
\over2lm}.
\end{equation}
Comparing to the previous section, we see that we can crudely think
of a planar Feynman diagram as a (coarsely) discretized world sheet,
with a dynamical link dependent string tension $T_{lk}=(l/k)T_0$.
Each link has its own independent $k,l$ which are each summed over all
positive integers. Of course only the large fishnet diagrams will
bear any actual resemblance to a continuous world sheet! The sum over
all planar diagrams would then define the QCD string dynamics as
including an average over all such discretizations ranging from
coarse to fine. Also note that a good
world sheet path integral should have (effectively) only 
positive weights. For instance, for $\lambda\phi^4$ theory each
vertex contributes a minus sign if $\lambda>0$. A good world sheet
interpretation requires $\lambda<0$, the attractive unstable sign.
Gauge theories have vertices of both signs, complicating
a straightforward world sheet interpretation. 

\begin{figure}[ht]
\centerline{\psfig{file=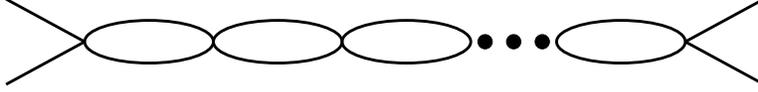,width=4.0in}}
\caption[]{Diagrams of $\phi^4$ scalar field theory summed in the text. }
\label{twototwo}
\end{figure}

To illustrate the effect of our discretization on a standard 
diagrammatic calculation, consider the sum of those 2 to 2 
scattering diagrams in $\lambda\phi^4$ theory shown in Fig. \ref{twototwo}. 
We work in the transverse center of mass frame. 
Assume that
the (discrete) $P^+/m$ of the initial (final) particles is $l, M-l$
($r, M-r$). Let $E=M{\bf p}^2/2l(M-l)$ be the initial energy ($P^-$).
Fix the discrete time of the first vertex at $0$, and sum over
all diagrams in which the final particles both propagate
to time $k$, multiply by $e^{kaE}$ and sum over all $k$ from
1 to $\infty$. The result for the off-energy shell S-matrix is then
\begin{eqnarray}
S_{fi}&=&\delta_{rl}\delta({\bf p}-{\bf p}^\prime)+{-\lambda\over
32\pi^3\sqrt{l(M-l)r(M-r)}}\nonumber\\
&&\left(e^{-aE+M{\bf p}^{\prime2}/2r(M-r)T_0}-1\right)^{-1}
\left(1-{\lambda(M-1)\over16\pi^2M}\ln(1-e^{aE})\right)^{-1}
\end{eqnarray}
Compared to standard formal scattering theory, we see that instead of
a factor $1/(E_f-E-i\epsilon)$ which acts in wavepackets like
the standard energy conserving delta function, the use of discrete
time has rendered this as $1/(e^{aE_f-aE}-1)$. This replacement is easy
to understand because with discrete imaginary time, the amplitudes
should be periodic in $E$ with period $2\pi i/a$. It is thus
apparent that the scattering amplitude should be identified with the
coefficient of this factor. For $\lambda<0$, the scattering amplitude
shows a bound state pole at a real negative value of $E$:
$$E_B\equiv -B={1\over a}\ln\left[1-e^{16\pi^2M/\lambda(M-1)}\right].$$
In the continuum limit, $M\to\infty,a\to0$, one can make the
pole location stay finite by tuning $\lambda$ to vanish logarithmically
as $a\to0$, showing dimensional transmutation in an asymptotically
free theory.
\section{Discretization Setup for Yang-Mills Field Theory}
\label{discreteym}
\setcounter{equation}{0}
The discretization of QCD initially attempted 
by Brower, Giles, and me \cite{browergt},
was based on a literal transcription of the 
Feynman rules in light-cone gauge. The transverse gauge field
can be described in the complex basis $(A_1\pm iA_2)$ when it
takes on the guise of a complex scalar field, described diagrammatically
by attaching an arrow to each line of a Feynman diagram. The
primitive quartic vertex conserves arrows, but the cubic vertices
can act as sources or sinks of arrows. The longitudinal gauge field $A_+$
does not propagate and can be integrated out to yield
an induced quartic vertex, which depends upon the
$P^+$ values of the incoming legs in a singular way:
\begin{equation}
\Gamma^4_{\rm induced}=g^2{(P^+_1+P^{+\prime}_1)(P^+_2+P^{+\prime}_2)\over
(P^+_1-P^{+\prime}_1)^2}.
\end{equation}
Upon discretization, we adopted the drastic prescription
of simply dropping the infinite contribution at $P^{+\prime}_1=P^+_1$.
Furthermore all tadpole diagrams had to be dropped, because
of our insistence that no line propagate 0 time steps.
Then the strong coupling limit singled out large planar diagrams
involving only the primitive quartic couplings, thus leading
to an evaluation similar to the $\phi^4$ example of the previous section.
Unfortunately, these quartic couplings have mixed signs: an
attractive interaction between gluons of parallel spin and repulsive
between gluons of antiparallel spin. This ferromagnetic
interaction pattern meant that our discretization
led to a formal strong coupling limit in which the only 
long string that could form in the limit would have huge
total spin. The essential problem is that attractive
interactions between gluons of opposite spin arise in QCD
from gluon exchange \cite{thorngluebag}, and the discretization chosen in
\cite{browergt} prevents anti-ferromagnetic gluon exchange from
competing at strong coupling with the ferromagnetic quartic interaction.   
This, together with our drastic prescription
for all of the $P^+=0$ ills of light-cone quantization, points
to the need for a more refined discretized model to adequately describe 
the strong coupling behavior of large $N_c$ QCD.
\subsection{Improved Discretization Rules}
\begin{figure}[ht]
\centerline{\psfig{file=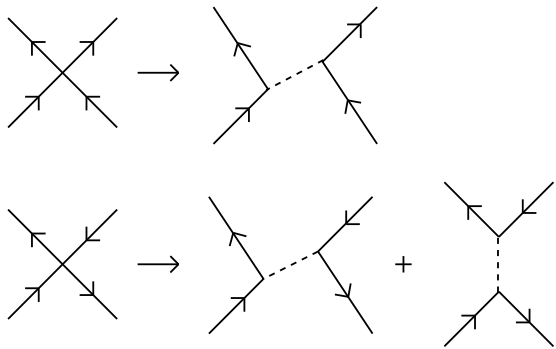,height=1.8in}}
\caption[]{Quartic vertices get replaced by two cubic vertices and a
fictitious scalar field.}
\label{4VertexTo3Vertex}
\end{figure}

\begin{figure}[ht]
\centerline{\psfig{file=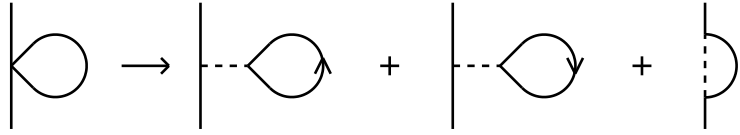,height=0.7in}}
\caption[]{Three tadpole diagrams resulting from the spreading out of
the quartic gauge vertex.}
\label{Tadpole2}
\end{figure}

\begin{figure}[ht]
\begin{center}
\begin{tabular}{|c|c||c|c|}
\hline
&&&\\[-.3cm]
$
\begin{array}[c]{c}
\psfig{file=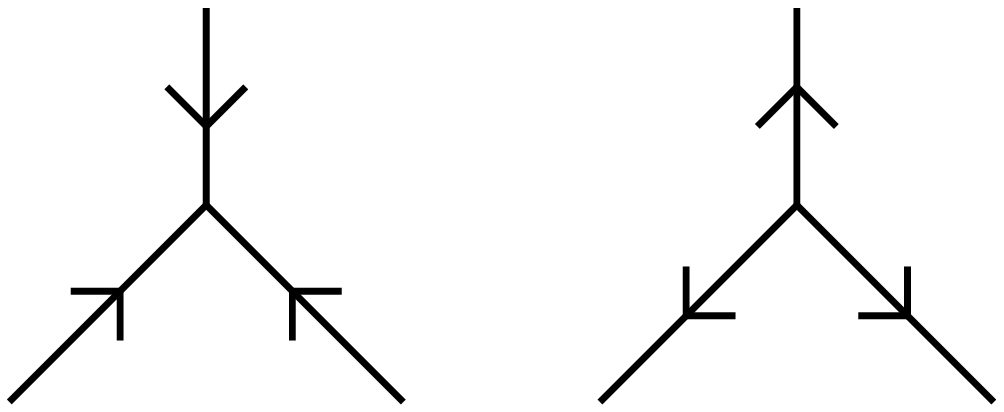,height=0.5in}
\end{array}
$
& 
$0$
&
$
\begin{array}[c]{c}
\psfig{file=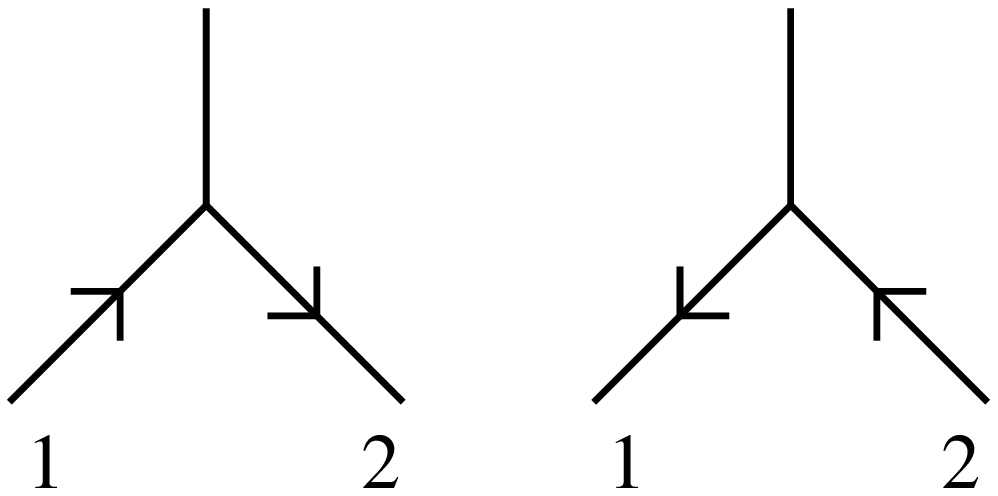,height=0.6in}
\end{array}
$
& 
$ {g\over T_0}(M_2-M_1)$ \\
\hline
&&&\\[-.3cm]
$
\begin{array}[c]{c}
\psfig{file=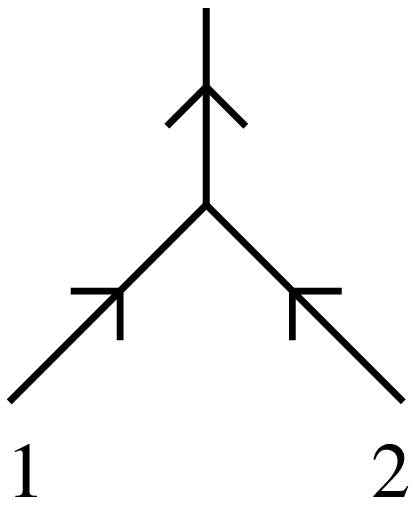,height=0.6in}
\end{array}
$
& 
$ -2{g\over T_0} \left({M_1+M_2\over M_1M_2}\right)
(M_1Q^\wedge_2-M_2Q^\wedge_1) $
&
$
\begin{array}[c]{c}
\psfig{file=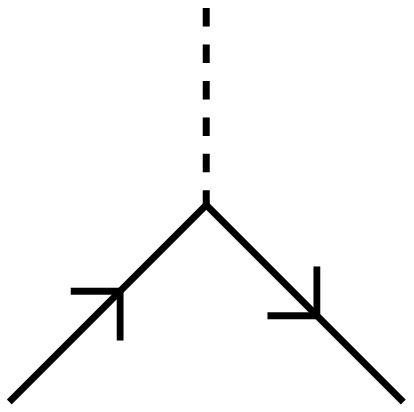,height=0.6in}
\end{array}
$
& 
$+{g\over T_0}$ \\
\hline
&&&\\[-.3cm]
$
\begin{array}[c]{c}
\psfig{file=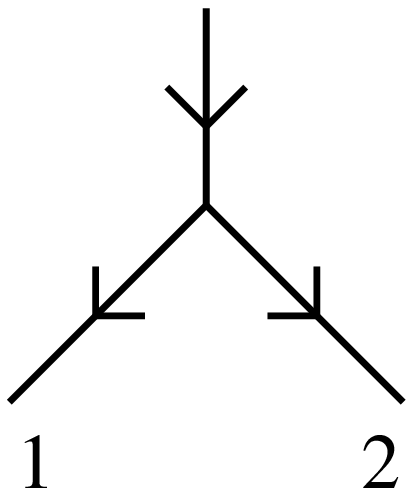,height=0.6in}
\end{array}
$
& 
$ -2{g\over T_0} \left({M_1+M_2\over M_1M_2}\right)
(M_1Q^\vee_2-M_2Q^\vee_1)$
&
$
\begin{array}[c]{c}
\psfig{file=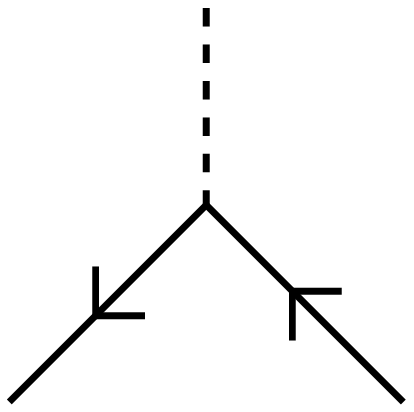,height=0.6in}
\end{array}
$
& 
$ -{g\over T_0}$ \\
\hline
&\multicolumn{3}{c|}{}\\[-.2cm]
$
\begin{array}[c]{c}
\psfig{file=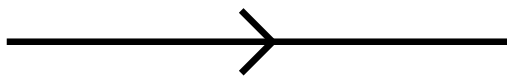,width=0.6in}
\end{array}
$
&
\multicolumn{3}{c|}{
${1 \over 2M} e^{-k{\bf Q}^2/2MT_0}$
} \\[.3cm] 
\hline
&\multicolumn{3}{c|}{}\\[-.2cm]
$
\begin{array}[c]{c}
\psfig{file=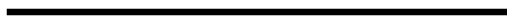,width=0.6in}
\end{array}
$
&
\multicolumn{3}{c|}{
$-f_k {T_0\over M^{2}}e^{-k{\bf Q}^2/2MT_0}$
} \\[.3cm] 
\hline
&\multicolumn{3}{c|}{}\\[-.2cm]
$
\begin{array}[c]{c}
\psfig{file=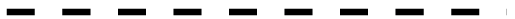,width=0.6in}
\end{array}
$
&
\multicolumn{3}{c|}{
$-h_k T_0 e^{-k{\bf Q}^2/2MT_0}$
} \\[.3cm] 
\hline
\end{tabular}
\end{center}
\caption[]{Summary of discretized Feynman rules using only cubic
vertices. We have explicitly inserted a factor of $1/T_0$ for
each vertex arising from the discretization.}
\label{NewRules}
\end{figure}

Clearly what is needed is a prescription that either enhances
strong coupling gluon exchange or suppresses the strong
coupling quartic interactions. We found it most natural to arrange
the latter by abolishing all quartic interactions, primitive
and induced, and replacing them with the exchange of
short lived fictitious particles. This is shown for the
primitive quartic interaction in  Fig.\ref{4VertexTo3Vertex}.
The dashed line is associated with the fictitious two-form propagator
\begin{equation}
\Delta_{\rm2-form}=-h_ke^{-k{\bf Q}^2/2lT_0},\qquad\qquad
\sum_{k=1}^\infty h_k=1,
\end{equation}
where the $h_k$ are tunable parameters which are required to vanish
rapidly with $k=1,2,\ldots$, the number of time steps propagated.
We treat the induced quartic interaction in a similar fashion,
giving the non-dynamical field $A_+$ a short-time propagator
\begin{equation}
\Delta_{+}=-f_ke^{-k{\bf Q}^2/2lT_0},\qquad\qquad
\sum_{k=1}^\infty f_k=1.
\end{equation}
The presence of the tunable parameters
$f_k$ and $h_k$ is very welcome, because they can be adjusted
to arrange the cancelation of cut-off artifacts that can typically
spoil Poincar\'e invariance in the continuum limit.
As a bonus, we find that our prescription provides the
appealing interpretation of tadpole diagrams indicated in
Fig. \ref{Tadpole2}.
The first two diagrams cancel exactly (which is fortunate since
they can't be drawn in our discrete model!) 
leaving the third diagram which {\it can}
be drawn. Our complete set of Feynman rules is neatly summarized in 
Fig. \ref{NewRules} taken from our paper \cite{beringrt}.
Note that to avoid clutter we have suppressed the
double line notation so these rules are completely sufficient 
for all graphs of planar or cylindrical
topology ($N_c\to\infty$). For general diagrams, the double line notation must
be restored in order to properly account for the $1/N_c$ corrections.

An easy way to understand why a set of rules with only cubic vertices
is possible, is to apply light-cone gauge to the Yang-Mills Lagrangian
in first order form. The upshot is the Lagrangian
\begin{eqnarray}
{\cal L}&=&{1\over2}\Tr\partial A_k\cdot\partial A_k+
{1\over2}\Tr{\hat A}^2+
\Tr\phi_{kl}^2
-2ig\Tr{1\over\partial_-}A_k[\partial_-A_n\partial_k A_n
-\partial_k A_n\partial_-A_n]\nonumber\\
&&-ig\Tr{1\over\partial_-}{\hat A}[A_k,\partial_-A_k]
-ig\Tr\phi_{kn}[A_k,A_n],
\end{eqnarray}
where ${\hat A}\equiv\partial_-A_+-{\bf\nabla}\cdot{\bf A}$. This
makes it clear why we called the fictitious scalar represented by the
dashed lines a 2-form: it is a (pseudo) scalar only in 3 + 1 dimensions.
\subsection{One Loop Self Energy}
Quite apart from its use as a facilitator for a strong
coupling expansion, our discretization can also serve as
a novel way to regulate the diagrams of weak coupling
perturbation theory. To illustrate this aspect, we quote the
result for the one loop gluon self-energy, with discretization
in place:
\begin{eqnarray}
{\Pi}_2(Q^2) 
&=&{g^2N_c\over4\pi^2}
\bigg[\sum_{k=1}^\infty{u^k\over k^2}
\left(4M[\psi(M)+\gamma]-{(M-1)(11M-1)\over3M}
\right)\nonumber\\
& &
-\sum_{k=1}^\infty u^k\sum_{l=1}^{M-1}{lh_k(l)\over Mk}
-\sum_{k=1}^\infty f_k{u^k\over k}
\left(4M[\psi(M)+\gamma]-{7(M-1)\over2}\right)
\bigg],
\label{pi2new}
\end{eqnarray}
where $u=e^{-Q^2/2MT_0}$. In order to cancel lattice artifacts
in the continuum limit $M\to\infty$,
we find the constraints
\begin{eqnarray}
\sum_{k=1}^\infty{f_k\over k}={\pi^2\over6},\qquad\qquad
\sum_{k=1}^\infty {h_k(l)\over k}=-{\pi^2\over18}\left(1-{1\over l}\right).
\end{eqnarray}
Then we obtain
\begin{eqnarray}
\Pi_2\to{g^2N_c\over16\pi^2}{Q^2\over T_0}\left\{
\left[8(\ln M+\gamma)-{22\over3}\right]\ln{Q^2\over2MT_0}+{4\over3}
\right\}.
\end{eqnarray}
Here $2MT_0$ functions as a uv cut-off. With this understanding
our result for $\Pi_2$ agrees exactly with the known light-cone gauge
result \cite{thornfreedom}. Notice that the parameters $f_k,h_k$
which specify our discretization enter weak coupling physics only
through their {\it moments}, for example $\sum_k  f_k/k$. In contrast,
the strong coupling limit is sensitive to the values of these parameters 
at low $k$. Thus the two limits give complementary constraints
on these parameters.
\section{Concluding Remarks}
\label{summation}
\setcounter{equation}{0}
Our main aim in developing this discretization formalism is
to establish a framework for handling the sum of all the
planar diagrams of $N_c\to\infty$ gauge theories. As yet we don't
have any dramatic results to report. However we have begun studying
how the machinery works in simpler situations than full-blown
gauge theories. In our paper \cite{beringrt} we worked out the sum of the
densest (strong coupling ``fishnet'') planar diagrams of $\Tr\phi^3$
scalar field theory. As expected
the result leads to the light-cone quantized bosonic string
(with all of its usual pathologies, including the tachyon).
The presence of tachyons is not particularly surprising, since
the energy density of the theory is unbounded below.
We have not made analogous progress on the corresponding diagrams of
gauge theories. But we have studied the latter theory in
the sectors with $M=2$. This is not particularly difficult, since
the limitation on $M$ reduces the sum of all possible diagrams
to a geometric series. Nevertheless, it is interesting, for example,
that the $M=2$ gluon propagator summed to all orders in perturbation 
theory displays no additional poles beyond that of the
massless gluon itself. In contrast the $M=2$ propagator for the
fictitious 2-form field shows a bound state pole at sufficiently strong
coupling $g^2N_c/8\pi^2>18.28$. This indicates that the 2-form
field may be particularly 
important for the understanding of the strong coupling
limit.

Another relatively simple testing ground for our formalism is
quantum field theory 
in low space-time dimensions, the simplest being the 't Hooft
model (QCD in one space and one time dimensions). Rozowsky
and I are just finishing up a study of this model. Since
this model is well understood even in the continuum limit,
we used it mainly as a test of how our model approaches
the continuum theory. One interesting feature of our
simultaneous discretization of $P^+$ and $x^+$ is that
one can approach the continuum in different directions.
For example the approach to continuum at fixed $T_0\equiv m/a$
is different from the approach studied in conventional DLCQ.
The latter keep $x^+$ continuous throughout (in our language
this means taking $a\to0$ {\it first} followed by $m\to0$). We have
confirmed that the same continuum  physics emerges in
both cases.

I would like to conclude this talk with some remarks on
longer term prospects and goals. Our renewed efforts to
sum planar graphs have been directly stimulated by the
ADS/CFT duality \cite{maldacena} proposed in the last couple of years.
This duality in turn was recognized \cite{wittenholog,gubserkp} 
to be a higher dimensional realization of 't Hooft's concept
of holography \cite{thoofthol}: the vision that a 
consistent quantum theory of gravity requires our
apparently 3 dimensional spatial world to be 2 dimensional.
I have advocated string bits as
a way to realize holography in 't Hooft's original 2 dimensional sense:
the two dimensions being the transverse dimensions of light-cone string.
However, the QCD gluons of this talk really live in 3 dimensions
in spite of their description on the light-cone: the longitudinal
dimension hasn't disappeared. Rather,  it has been disguised as a variable
Newtonian mass. (In the ADS/CFT duality this third dimension
gets interpreted as a fifth dimension, whence holography is the 
statement that a 4+1 dimensional effective theory arises from a
3+1 dimensional quantum field theory.)  The defining character
of string bits  is that they have a {\it fixed} Newtonian mass,
in sharp contrast to the gluons we have been describing. To 
understand 3+1 gauge theories as part of
a string bit theory, a gluon with $P^+=lm$ must in reality be
a composite of $l$ string bits: the gluon vertices would
then be effective fission/fusion amplitudes as in
nuclear physics, rather than fundamental interactions.

\noindent\underline{Acknowledgments:} Most of the work described
in this talk was done in collaboration with Klaus Bering and Joel
Rozowsky, whom I thank for their essential contributions and
insights.


\end{document}